INDCOR

INDCOR WHITEPAPER 2

# Interactive Narrative Design for Representing Complexity

EDITORS

Andrew Perkis, Mattia Bellini, Valentina Nisi and Shafaq Irshad

AUTHORS

Andrew Perkis, Mattia Bellini, Valentina Nisi, Maria Cecilia Reyes, Cristina Sylla, Mijalche Santa, Anna Zaluczkowska, Shafaq Irshad, Ágnes Karolina Bakk, Fanny Barnabé, Daniel Barnard, Nadia Boukhelifa, Øyvind Sørdal Klungre, Hartmut Koenitz, Vincenzo Lombardo, Mirjam Palosaari Elahdhari, Catia Prandi, Scott Rettberg, Anca Serbanescu, Sonia Sousa, Petros Stefaneas, Dimitar Uzunov, Mirjam Vosmeer, Marcin Wardaszko

INDCOR

This whitepaper has been produced by members of the COST Action 18230 – Interactive Narrative Design for Complexity Representation (INDCOR - www.indcor.eu)

**EDITORS**
Andrew Perkis, Mattia Bellini, Valentina Nisi and Shafaq Irshad

**AUTHORS**
Andrew Perkis, Mattia Bellini, Valentina Nisi, Maria Cecilia Reyes, Cristina Sylla, Mijalche Santa, Anna Zaluczkowska, Shafaq Irshad, Ágnes Karolina Bakk, Fanny Barnabé, Daniel Barnard, Nadia Boukhelifa, Øyvind Sørdal Klungre, Hartmut Koenitz, Vincenzo Lombardo, Mirjam Palosaari Elahdhari, Catia Prandi, Scott Rettberg, Anca Serbanescu, Sonia Sousa, Petros Stefaneas, Dimitar Uzunov, Mirjam Vosmeer, Marcin Wardaszko

INDCOR WHITEPAPER 2

# Interactive Narrative Design for Representing Complexity

This article/publication is based upon work from COST Action INDCOR, CA18230, supported by COST (European Cooperation in Science and Technology).

COST (European Cooperation in Science and Technology) is a funding agency for research and innovation networks. Our Actions help connect research initiatives across Europe and enable scientists to grow their ideas by sharing them with their peers. This boosts their research, career and innovation.

www.cost.eu

COST EUROPEAN COOPERATION IN SCIENCE & TECHNOLOGY

Funded by the European Union

# Table of Contents

# Executive Overview

This white paper was written by the members of the Work Group focusing on design practices of the COST Action 18230 – Interactive Narrative Design for Complexity Representation (INDCOR, WG1). It presents an overview of Interactive Digital Narratives (IDNs) design for complexity representations through IDN workflows and methodologies, IDN authoring tools and applications. It provides definitions of the central elements of the IDN alongside its best practices, designs and methods. Finally, it describes complexity as a feature of IDN, with related examples.

In summary, this white paper serves as an orienting map for the field of IDN design, understanding where we are in the contemporary panorama while charting the grounds of their promising futures.

# Introduction

Since the world around us is becoming increasingly complex, we need new ways to represent it. Interactive Digital Narratives (IDNs) offer a unique form of representation due to the specific characteristics of the digital medium. IDN can present different perspectives within the same work, show the results of different decisions, and allow audiences to replay and make different choices (see INDCOR WP 0, 2023). Most importantly, IDNs transform audiences from passive readers to active participants, allowing them to experience complexity first-hand instead of merely reading about it (the so-called interactors - see below).

The potential of IDN for representing complexity has already been realized in the forms of video games and interactive documentaries, as well as in hypertext fiction, installation pieces and XR experiences. So far, however, IDN design remains highly idiosyncratic and is often not well understood outside the circle of successful IDN creators and scholars who specialize in the field.

This white paper addresses the situation by providing an accessible overview of IDN design for complexity representations. The discussion will focus on the power of the digital medium and its unique feature of interactivity, which define these cultural and often artistic works. Interactivity sets IDN apart from traditional narratives and will be a central topic in this paper.

A wide range of topics related to IDN design will be presented in the following sections, including IDN workflow and methodology, IDN authoring tools, and IDN application. To guide the reader without disrupting the flow of the text, we have provided definitions of the central elements of the practice in boxes on the side of the main text.

The main goal of this white paper is to serve as an informative guide to the field of IDN design, providing readers with a clear understanding of the current state of the field and highlighting the latest developments and promising future directions. By charting these aspects, this white paper will offer an orienting map for those who wish to explore this exciting area of digital media.

INDCOR

# 01. Definitions

This section provides our definition of Interactive Digital Narrative (IDN) and some of the key elements IDN is built on. In INDCOR, we understand IDN as follows:

IDN are narrative experiences that can be changed by an audience and which are created for the digital medium. What can be changed (for example outcome, progression, perspective) varies as does the particular form. Popular IDN forms include narrative-focused video games, interactive documentaries, hypertext fiction and VR/AR/MR experiences.

The table below defines some of the central elements of the practice.

| | |
|---|---|
| **Digital** | All objects, including media and text that exist in digital form and that can be read by machines. |
| **Narrative** | An account of a real or fictional sequence of events that may be presented in different ways (text, speech, images, animations, videos, or any combination of these). Narratives are ways to structure and transfer information (like knowledge, feelings, experiences, etc.) shared by all human cultures. |
| **Interactivity** | Describes an active relationship between two or more entities, people or objects. In digital media, interactivity represents a two-way flow of information between the devices and its user. In other words, it is the ability of a computer to respond to the user's input. |
| **Complexity** | The phenomenon of having many parts that cooperate in a sort of symbiosis to give rise to a whole that is constantly evolving, and that is more than the sum of its parts. |
| **Representation of complexity** | The process or product of depicting a complex phenomenon, real or fictional. In the context of this text, these depictions (or representations) are meant to make such phenomena more understandable. |
| **IDN authoring** | A creative process that comprises all the tasks that are involved in the creation of an IDN artifact, from ideation to distribution. An IDN author is a person who participates in the ideation, planning, design and/or development of an IDN. |
| **Authoring tool** | A software program that assists or facilitates the creation of an IDN artifact. By providing a unified and comprehensive workspace, it simplifies the authoring process. |
| **Interactor** | The user of an IDN. The term is meant to highlight the active participation enabled by IDNs, in contrast to the more passive consumption of representations in novels or films. |
| **Metaverse** | An online virtual world that multiple people can access at the same time. The currently available version of the metaverse has simple graphics, it's interactive and allows the creation of avatars to interact with other users. |
| **Virtual Reality (VR)** | A computer-generated simulation of a three-dimensional space that a person can interact with using special equipment (like headsets, controllers, gloves, etc.). It is a type of Extended Reality. |
| **Extended Reality** | A broad term referring to various combinations of real and virtual environments, as well as human-machine interactions generated by computer technology and equipment. It includes a number of specific forms like augmented reality, mixed reality and virtual reality. |

# 02. Design and Best practice for creating IDNs

This section explains how to create an IDN: what are the best design practices, what tools are available for creating IDNs, and how would one go about planning and making an IDN. There are several ideation, prototyping, and planning methods that can be used, from informal brainstorming to more structured approaches.

While the specific process of creating an IDN is unique to each work, there are some common practices that can be highlighted. As for most design tasks, it is common to work iteratively and undertake user testing. IDNs are sometimes created in collaboration with their users, who make an active contribution during the design, just as in standard user-centered approaches. There exist many tools for creating IDNs, each designed to cater to different types of artifacts the creators might want to produce.

An authoring process for IDNs generally has four main stages: ideation, pre-production, production, and post-production, with specific tasks for each stage that can be adapted to the particular IDN project. As noted in the INDCOR WP 0 (2023) there are no established production practices as well as production facilities; so, what we report here is an abstraction on singular experiences. It is important to note that the process of creation is not linear, and often involves going from ideation to production and back again: the four-stage model aims to be inclusive and iterative. It is open to the addition of new actions into each stage and to the repetition of actions to accomplish the desired results.

1. Ideation methods and early prototyping
   a. The ideation phase of an IDN is hard to abstract, as every author of any artifact proceeds in different ways. Some authors might find it useful to start by developing a range of ideas for the broad structure of the IDN and picking the best one (considering factors such as expressiveness, quality of the experience, feasibility, and production costs). Other authors may start with clear ideas for the IDN content and draft an early prototype from there. The ideation stage also comprises research and training work on the thematic of the IDN, or on the selected tools and materials that the IDN will require. During this stage, authors can also sketch an overall interactive structure with possible plot(s), create characters, draft events, and make notes for the next stage.
2. Pre-production and interaction design
   a. A key feature of IDNs is their interactivity: the interactor is active in the experience. A central question is therefore, "What is the role of the interactor?" or simply "what can the interactor do?" Decisions about the role of the interactor affect the dynamics and mechanics of the experience, the meaning ascribed to the experience, and the emotion it elicits. One approach to the early creation stages is to list different options for the role of the interactor(s) and then brainstorm what opportunities and challenges each would create, both for the interactor and for the author.
3. Production – an iterative process
   a. The process of creating IDNs is usually iterative, i.e. a version is created, tested, improved and tested again until its authors are satisfied with it. This is particularly important for IDNs, as their interactivity makes more difficult to predict the actions and responses of the interactors. This iterative testing frequently narrows down to smaller and smaller sets of components, with early iterations modifying the main elements of the IDN and later iterations concentrating on finer details, such as locating software bugs or even the precise wording of instructions.
4. Post-production and testing
   a. Informal testing usually happens throughout the whole design process. In addition, it might be helpful to perform a more formal evaluation towards the end. This can be done by having a group of people test the IDN and obtain their feedback through interviews and/or questionnaires. At this stage, the purpose is not necessarily to fix or improve the IDN at hand. Instead, it could be helpful to get information for future design processes. In a research context, the goal could also be to test a hypothesis by comparing different variations of the IDN.

# 03. IDN Authoring tools

INDCOR WG1 has been actively working on compiling an archive[1] of existing IDN authoring tools, organizing them according to taxonomic criteria and design principles, and identifying potential testing methodologies.An IDN authoring tool is pragmatically defined as digital software that eases the authoring process of interactive digital narratives. In particular, an IDN authoring tool:

1. Provides an independent and comprehensive workspace, including an IDE and GUI, which allows a prospective author to create an IDN work from start to finish. The tool gears its end products to run on an engine, which is often, though not necessarily, also embedded within the tool's environment.
2. Simplifies the authoring process: the tool streamlines the design of the story world, the end-user interaction model, and/or other central narrative elements such as characters and events, to make the IDN creative process easier and more effective than design through a general-purpose programming language would be.
3. Is actively being used or was actively used in the past to create IDN products, focused on interactive narrative aspects, by a community of practitioners besides the tool's creator(s). Prototypes and in-house tools may also meet this criteria, if they demonstrate clear and explicit potential to be actively used in the future, when they are publicly released.

# 04. Complexity IDNs

This section will focus on a specific topic on which the COST Action INDCOR has focused significant attention in the last few years, namely the relation between complexity and IDNs. In our understanding, complexity represents a significant societal challenge, particularly when materializing through contemporary complex phenomena such as global warming, refugee crises, wars, colonization, globalization, pandemics and so forth (see INDCOR WP 0, 2023).

In this section, two interpretations of the relation between IDNs and complexity will be touched on, namely complexity as a feature of IDNs (and what it means for authors), and the suitability of IDNs to better represent complex topics and phenomena.

# 05. Complexity as a feature of IDNs

In IDN design, complexity can be intended as a feature of the object in question. The authors decide on the level of complexity of a particular IDN through a number of choices regarding game rules and narrative characteristics. As a feature of IDNs, complexity can be manually crafted by authors (for example, by building extremely large worlds with many interacting components) or it can sprout from the engagement of interactors with the IDN (as in the game of chess, in which authors defined very few elements but with a huge number of possible moves at each step). Complexity as a feature is necessary to provide a meaningful interaction, as it relates the interactors' choices to their outcomes, making IDNs ideal candidates for representing complex issues in an understandable way. For example, an IDN on the migration crises in the Mediterranean Sea or piracy in the Indian Ocean (see INDCOR WP 0, 2023) needs to present a number of components that revolve around the issue (people, countries, governments, policies, NGOs, etc.), and needs to allow a number of possible interactions to make this system understandable.

However, complexity as a feature also requires precautions. First of all, while designing IDNs with elements of complexity, authors must remember that interactors face the opposite end of their creation: what can look simple for the author might be complex for the audience and vice versa. A second caveat is that often complexity as a feature has to be presented to interactors gradually, to get them acquainted with this complexity step-by-step, and not overwhelm them, which might prevent them to understand what is represented and why. On the other hand, it can be sometimes useful to deliberately avoid this gradual presentation to achieve specific expressive purposes, like representing the complex situations some people face in their daily lives.

[1] https://omeka-s.indcor.eu/s/idn-authoring-tools/page/idn-authoring-tools

# 06. IDNs representing complexity

IDNs have significant advantages compared to traditional narratives in representing complexity: they can represent a topic with a great wealth of details, they feature dynamic representations that can change and adapt in response to user choices, they can accommodate a number of even conflicting perspectives or points of view, and they allow multiple replays to help understanding. The same kind of representation is not possible to achieve (for instance) in the text-based newspaper article, nor in the fixed form of a TV broadcast (see INDCOR WP 0, 2023). Even when talking about fiction, readers/viewers of traditional forms like novels or movies can only speculate about the progression and outcome of the story, but they cannot influence them in any way. In IDN, users can make plans, execute them and see the results of their actions.

Choices and their resulting consequences, as well as re-play, is what sets IDNs a separate entity from fixed narratives. These qualities are necessary for enabling an understanding of the dynamic nature of complex issues, allowing them to represent systems in a more direct way, and permitting users to practically engage with such systems and understand their functioning. This capability is crucial to train systemic thinking, which is the ability to understand the dynamic and interconnected nature of all phenomena in contemporary society. In addition, by understanding and engaging with these complex systems, even though fictional, users can be prompted to change their behavior regarding complex matters in real life, for example when deciding their political position, or they can change their perspective and point of view, for example by crossing the boundaries of anthropocentrism and adopting a multi-species point of view as a way to restore equilibrium among the inhabitants of our planet.

Representing complex issues through IDNs requires specific design choices and precautions:

- The wealth of details should be presented in a way that is not overwhelming for the user and that is easily understandable: an IDN is not an encyclopedia, and the information presented to the users should be selected on the basis of relevance for the scope of the object being designed, and of manageability (also in terms of practical and economic feasibility). The example IDN representing the issues connected with the migration crises in the Mediterranean Sea does not need to present all technical details of the rescuing ships, but only the relevant ones (e.g. load capacity, speed, etc.).
- The IDN as a dynamic representation, too, sprouts from design choices: the degree of adaptability of the IDN system should come from choices made early in the design process, as they can greatly affect the dimensions of the final object and therefore its feasibility. Again, this is a matter of finding equilibrium between fidelity of the object to the issue it is intended to represent, comprehension of the outcome, technical, practical and economic feasibility, and accuracy of the representation of the topics in question. Our IDN on migration crises could allow users to choose between different immigration policies to see the long-term effects of them, while other kinds of choices on social issues could be disregarded as they might have little impact on the subject, like the retirement age of involved countries.
- The conflicting perspectives should be balanced so that their respective weights allow for sufficient freedom of exploration: which also allows for freedom of choice and is eventually the prerequisite of systemic thinking. If only one perspective is represented or is dominant, IDN's ability to make a complex issue understandable will diminish. If the IDN on the migration crisis represents only or mainly the point of view of the extreme right, it would easily miss its scope of representing the complexity of the issue. This also highlights a possible ethical dilemma connected with IDN, namely the misuse or even biased use of them to support political or ideological agendas while pretending to be impartial and mindful of the different perspectives.
- Finally, the re-play should be encouraged by the design: re-plays are great enhancers of systemic understanding, as they allow users to achieve the multiperspectivity discussed in the previous point and greatly enhance their understanding of the whole system, by observing its behavior in different conditions. While users can always restart interacting with an IDN, this re-engagement can be encouraged and can be made more relevant through specific design choices. For example, by including maps or diagrams that also show not-yet-seen paths, or by hinting at the presence of multiple perspectives and endings after having explored one of them.

These features make IDNs ideal candidates to address complexity as a societal challenge when materializing in complex phenomena and issues. However, more particular application areas of IDN design for representing complexity could be highlighted.

# 07. Application areas

This section aims at providing an overview on IDNs that address complex issues to gain a better understanding of the design decisions that have been made in those prototypes and applications, and then to make the resulting knowledge available to practitioners for the representation of complex topics. This section corresponds well to the start of INDCOR WG1, where we made an inventory of available applications of IDNs in various domains.

When applied to representing complex issues, IDN design finds some important application domains, such as:

- Journalism and news
  - ◊ Journalism has benefited much from the digital medium. From interactive versions of magazines and newspapers, where readers can comment and add their voice to the authorial one, to blogs, Twitter and various types of social media which are transforming the panorama of journalistic practices. Moreover, at the cutting edge of virtual and extended reality, journalism becomes immersive with the pioneering work of the American journalist Nonny de la Peña. Efforts blossom worldwide as journalists, documentarists and filmmakers start challenging journalistic practice with IDNs. Interactive documentaries such as Last Hijack - Interactive, The Industry, or Miners Walk, The Criers of Medellin, and many others available at the MIT Documentary database (https://docubase.mit.edu) are just some of the examples (see INDCOR WP 0, 2023 for other examples).
- Education and training
  - ◊ Education is a vast area to apply IDN and its capability of representing complexity. "Edutainment", a word coming from the blending of "education" and "entertainment", has been identified as an area of development and research decades ago, with examples ranging from openly educational games for schools of all levels, to serious games, to interactive stories for children both individually and in classrooms. Examples of this application of IDNs are Periodic Fable, a story-game leveraging on characters and narratives to excite children in learning for instance chemistry subjects, Starlight Stadium, an IDN that teaches Human Rights monitoring practices, or Mobeybou, which requires young children to connect physical, tangible blocks representing characters, objects or places, to populate a virtual world where these elements interact in different ways, to ultimately teach the richness of distant cultures.

In addition, there are a number of other application domains being explored, such as entertainment and tourism. These, however, do not necessarily address the topic of complexity as defined by INDCOR.

Existing applications present a digital narrative within a specific domain. It is possible, however, to apply interactive digital narratives to more complex issues as technology advances. For example, explorations are made for the application of IDN to represent and enable understanding of complex topics at the general and individual levels. Climate change, wars, pandemics, migrations, traumas, other mental health issues, interpersonal relationships, etc., are examples of complex topics. As a result of this work, IDN can be used to represent, understand, and create solutions for wicked problems in the future. IDN can, for example, be used to explore solutions to social injustice, poverty and homelessness, pandemics, racism, globalization, poverty, homelessness, global warming, the current war in Ukraine in the future.

# 08. Future directions

### Augmented Reality, Virtual Reality, Extended Reality

With the focus on Extended Reality (XR), new exciting possibilities in the field of IDNs have manifested. First out was the recent revival of VR, which allows interactors to feel fully inside the action, and the design of IDNs for VR has to take the position of their interactors into account even more than when the narrative is presented through a flat screen. The ways of interacting with the narrative are other very important elements that designers have to consider: in VR, interaction can be induced with handheld devices, gestures or even gaze input. This means that an interactor can activate a reaction in the virtual environment just by performing a particular gesture or by looking at a certain point. While these features are mainly applied in simple ways that enable the interactor to control the progress of a linear story, the increasing affordability of these new technologies may allow future authors to explore fascinating new domains of IDN creation.

The combination of sensors, extended realities and AI opens up possibilities in uncharted territories for audiences and narrative creators. Narratives could be informed by real time data, processed directly from the interactors' senses, emotions, and context, shaping up and personalizing their experiences as they live through the narrative journeys designed for and with them. As technology starts to move its first steps in this direction, authors and audiences are challenged to embark on daring new narrative journeys and experiences.

However, an important caveat is to be pointed out: features that visionaries claim would be achievable in the near future, often seem to be based on technological fantasies (inspired by movies, books and television series), rather than on actual technical possibilities. It is not uncommon, for instance, to see predictions of online virtual worlds with unlimited interaction, photorealistic environments, and believable characters. With current technological affordances, however, most of these features are impossible, but these sensationalized expectations may cause users to easily become disappointed when they are confronted with versions of a Metaverse that are much more sober and constrained than the world that fictional stories have led them to imagine. Technological progress may actually evolve in a spectacular fashion that will enable users to be immersed in interactive movies that will be indistinguishable from the world around them. Until then, the affordances that online VR offers for interactive storytelling are still very much worth exploring.

## Artificial Intellingece

With digital narratives, we play the role of humans by following different storylines and creating open-ended stories. In the future, however, artificial intelligence and technological advancements will allow humans to no longer be the only agents responsible for creating storylines. We see that AI is being used in games to develop emergent behaviors for secondary non-player characters. It is likely that AI will become an essential component of interactive digital narratives in the future. Combining this with the domains and usage of IDN in complex issues and wicked problems, we can expect that artificial intelligence can have a more active role in creating interactive digital narratives to represent and understand complexity. Imagine, however, that AI is used in IDN to the fullest extent possible. It is possible, then, to imagine AIs as the only participants in an interactive digital narrative without human participants. So what will IDNs be used for if this happens?

The goal of interactive narrative is to immerse the user in an intellectual and emotional experience so that the user's actions can directly impact the direction or outcome of the storyline. Thus, the user is given the opportunity to "replay" the story in different forms. A replay will provide reflections on the various other, often opposing, directions and meanings the story might have, depending on the starting point and interactions. But, as IDN can be applied to more complex issues and wicked problems, there is an opportunity to enhance the purpose of IDNs towards a solution-finding role. The new IDN purpose may be to create solution storylines through co-creation between different stakeholders at various levels. In this way, IDN can be viewed as a method for addressing real problems that affect many people, such as those necessary for achieving the Sustainable Development Goals of the United Nations.

IDNs are slowly finding their space in a number of application areas, as we have seen (cf. also INDCOR WP 0, 2023). However, with the continuous development of new technologies, ever new directions and applications are just about to be opened.

## Extending IDNs to other contexts, stakeholders and audiences

IDNs are likely to be soon introduced in contexts that are traditionally considered less driven by digital narratives, such as data science and knowledge representation. It will be especially interesting to see the effects of their use in dealing with complex issues and multi-perspective situations. For example, the World Economic Forum developed what they call "transformation maps", interactive interfaces that "help users to explore and make sense of the complex and interlinked forces that are transforming economies, industries and global issues". Interactive digital narrative design could help to further enrich these increasingly popular tools by making use of the expressive power of narratives, a way to share information rooted in the very core of all human cultures, and ultimately make complex topics understandable. Throughout the text, we have highlighted practices for the design of interactive digital narratives, discussed why they can be ideal candidates for representing complexity and shown possible specific areas in which IDNs can represent meaningful and helpful choices. This, we hope, will help to address complexity as a societal challenge, to assist experts in reaching shared understandings, and to engage new audiences and stakeholders to eventually change the world for the better.

# Further Readings

Hargood, C., Millard, D. E., Mitchell, A., & Spierling, U. (Eds.). (2022). The Authoring Problem: Challenges in Supporting Authoring for Interactive Digital Narratives. Springer International Publishing. https://doi.org/10.1007/978-3-031-05214-9

Koenitz, H. (2023). Understanding interactive digital narrative: Immersive expressions for a complex time. Routledge.

Koenitz, H., Holloway-Attaway, L and Perkis, A. (2022) Frontiers special issue: Interactive Digital Narratives Representing Complexity

Nack, F. (2023). Interactive digital narrative (IDN)—a complexity case. New Review of Hypermedia and Multimedia, DOI: 10.1080/13614568.2023.2173385

Yotam Shibolet and Vincenzo Lombardo (2022). Resources for Comparative Analysis of {IDN} Authoring Tools, Proceedings of Interactive Storytelling - 15th International Conference on Interactive Digital Storytelling, ICIDS 2022, editors Mirjam Vosmeer and Lissa Holloway-Attaway, Santa Cruz, CA, USA, December 4-7, 2022, Lecture Notes in Computer Science 13762, pp. 513--528, Springer.

INDCOR WP 0 (2023) Interactive Digital Narratives (IDNs)– A solution to the challenge of representing complex issues, Koenitz, H., Barbara, J., Holloway-Attaway, L., Nack, F., Palosaari Eladhari, M., Bakk, A., https://arxiv.org/abs/2306.17498

INDCOR WP 1 (2023) A Shared Vocabulary for IDN. Koenitz, H., Palosaari Eladhari, M., Louchart, S., Nack, F., https://arxiv.org/abs/2010.10135

INDCOR WP 3 (2023) Interactive Digital Narratives and Interaction, Nack, F., Louchart, S., Lund, K., Bellini, M., Georgieva, I., Atmaja, P., W., Makai, P., https://arxiv.org/abs/2306.10547

INDCOR WP 4 (2023) Evaluation of interactive narrative design for complexity representation, Roth, C., Pitt, B., Šķestere, L., Barbara, J., Bakk, A., K., Dunlop, K., Grandio, M., Barreda, M., Sampatakou, D., Schlauch, M., http://arxiv.org/abs/2306.09817

INDCOR WP 5 (2023) Addressing social issues in interactive digital narratives, Silva, C., Gil, M., Holloway-Attaway, L., Aguado,J., M., Gërguri, D., Kazazi, L., Marklund, B., B., Zamora Medina, R., Fahmy, S., Noguera Vivo J., M., Bettocchi, E., Papaioannou, T., http://arxiv.org/abs/2306.09831

EUROPEAN COOPERATION IN SCIENCE & TECHNOLOGY

INDCOR

WWW.INDCOR.EU